\documentclass[9pt]{article}
\usepackage{spconf,amsmath,graphicx}
\usepackage{cite}
\usepackage{bibspacing}
\usepackage{booktabs}
\usepackage{color}
\usepackage{amsfonts}
\usepackage{multirow}
\usepackage{amssymb}
\usepackage{array}
\usepackage{authblk}
\usepackage{threeparttable}
\newcommand{\PreserveBackslash}[1]{\let\temp=\\#1\let\\=\temp}
\newcolumntype{C}[1]{>{\PreserveBackslash\centering}p{#1}}
\newcolumntype{R}[1]{>{\PreserveBackslash\raggedleft}p{#1}}
\newcolumntype{L}[1]{>{\PreserveBackslash\raggedright}p{#1}}
\usepackage{enumitem}
\usepackage{subfigure}
\usepackage[hang]{footmisc}
\usepackage{graphicx}
\usepackage[belowskip=-1pt]{caption}
\usepackage[skip=2.0pt]{caption}
\usepackage{bm}

\usepackage{times}
\usepackage{latexsym}
\usepackage{algorithm}
\usepackage{algorithmic}
\usepackage{multirow}
\usepackage{amsfonts}
\usepackage{amssymb,bm}
\usepackage[dvipsnames]{xcolor}
\usepackage[T1]{fontenc}
\usepackage{url}
\usepackage[utf8]{inputenc}
\usepackage{microtype}
\usepackage{inconsolata}

\usepackage{graphicx}
\usepackage{amsmath}
\usepackage{booktabs}

\usepackage{lipsum}

\AtBeginDocument{
 \footnotesize\setlength{\footnotemargin}{0pt}\normalsize
 \edef\hangfootparindent{\the\parindent}
 
}

\makeatletter
\newcommand\email[2][]%
   {\newaffiltrue\let\AB@blk@and\AB@pand
      \if\relax#1\relax\def\AB@note{\AB@thenote}\else\def\AB@note{\relax}%
        \setcounter{Maxaffil}{0}\fi
      \begingroup
        \let\protect\@unexpandable@protect
        \def\thanks{\protect\thanks}\def\footnote{\protect\footnote}%
        \@temptokena=\expandafter{\AB@authors}%
        {\def\\{\protect\\\protect\Affilfont}\xdef\AB@temp{#2}}%
         \xdef\AB@authors{\the\@temptokena\AB@las\AB@au@str
         \protect\\[\affilsep]\protect\Affilfont\AB@temp}%
         \gdef\AB@las{}\gdef\AB@au@str{}%
        {\def\\{, \ignorespaces}\xdef\AB@temp{#2}}%
        \@temptokena=\expandafter{\AB@affillist}%
        \xdef\AB@affillist{\the\@temptokena \AB@affilsep
          \AB@affilnote{}\protect\Affilfont\AB@temp}%
      \endgroup
       \let\AB@affilsep\AB@affilsepx
}
\makeatother

\title{S2ST-Omni: Hierarchical Language-Aware SpeechLLM Adaptation for Multilingual Speech-to-Speech Translation}

\author[1]{Yu Pan}
\author[2]{Xiongfei Wu}
\author[4]{Yuguang Yang}
\author[5]{Jixun Yao}
\author[3$\dagger$]{Lei Ma}
\author[1$\dagger$]{Jianjun Zhao}

\affil[1]{Kyushu University, Japan \quad $^2$University of Luxembourg, Luxembourg \quad $^3$The University of Tokyo, Japan}
\affil[1]{$^4$Tencent, China \quad $^5$ByteDance Ltd., China}

\email{panyu.ztj@gmail.com, ma.lei@acm.org, zhao@ait.kyushu-u.ac.jp}

\begin{document}
%
\maketitle

\begin{abstract}
Despite recent advances in speech-to-speech translation (S2ST), it remains difficult to achieve both high translation accuracy and practical flexibility. In this paper, we present \textit{S2ST-Omni}, a compositional S2ST framework that integrates a high-accuracy speech-to-text translation (S2TT) frontend with a modular, plug-and-play text-to-speech (TTS) backend, enabling independent optimization of translation and synthesis. On the S2TT side, we introduce a hybrid adapter that follows a "local-then-global" strategy to bridge a pretrained Whisper encoder and a Qwen3 LLM, yielding a hierarchical acoustic-to-semantic abstraction. Building on this bridge, we further propose a hierarchical language-aware architecture that injects source-language information at two complementary levels. At the acoustic level, \emph{Language-Aware Dual-CTC} operates on intermediate adapter features and employs FiLM-style feature modulation with a learnable gate, encouraging the model to learn language-specific but content-faithful acoustic representations. At the linguistic level, \emph{Language-Aware Prompting} dynamically constructs source-language-conditioned prompts that activate language-specific translation knowledge in the LLM. To enable efficient optimization, we design a task-specific progressive fine-tuning strategy that first stabilizes speech-text alignment and then improves translation via LoRA on top of this converged foundation. The TTS backend remains fully modular and can be instantiated with any state-of-the-art synthesizer without retraining the S2TT frontend. Experiments on CVSS-C show that \textit{S2ST-Omni} consistently achieves the best BLEU and ASR-BLEU across French, German, and Spanish to English directions, outperforming strong recent S2ST baselines.
\end{abstract}


{
\let\thefootnote\relax
\footnote{$\dagger$ denotes the corresponding author.}
}

\section{Introduction}
Multilingual speech-to-speech translation (S2ST) aims to directly convert utterances from multiple source languages into target language speech, which plays a pivotal role in cross-language communication, healthcare and education \cite{kikui2003creating,lee2022textless}.

Recent years have witnessed rapid progress in S2ST, with methods broadly falling into two families: end-to-end (E2E) and compositional. E2E approaches, exemplified by Translatotron series \cite{jia2019direct,jia2022translatotron} and SeamlessM4T \cite{barrault2023seamlessm4t}, map source speech to target speech within a single unified network. While conceptually attractive, E2E systems must jointly handle speech understanding, cross-lingual translation, and speech generation, 
which often leads to trade-offs that limit optimization of any individual sub-task. Moreover, the lack of an intermediate textual representation complicates error analysis and correction, and the scarcity of large-scale speech-speech parallel data further constrains scalability \cite{fang2024can}. 

To mitigate these issues, compositional S2ST architectures decompose the problem into an S2TT frontend and a TTS backend. For example, TranSpeech \cite{huang2023transpeech} operates in a discrete unit space and introduces bilateral perturbation for acoustic-style normalization, enabling non-autoregressive direct S2ST. DASpeech \cite{fang2023daspeech} adopts a two-pass design in which a linguistic decoder first predicts target-side representations and an acoustic decoder (FastSpeech~2 \cite{ren2020fastspeech}) then generates speech conditioned on them. These models improve controllability and latency, but are still trained as direct S2ST systems and heavily rely on costly source speech to target speech pairs. In contrast, ComSpeech \cite{fang2024can} proposes a genuinely compositional architecture that explicitly integrates pretrained S2TT and TTS models via CTC-based vocabulary adaptors, thereby leveraging abundant S2TT/TTS corpora and exposing an interpretable intermediate text. However, its vocabulary adaptors and phoneme-level embeddings yield relatively shallow, weakly contextualized interfaces, limiting robustness on semantically complex or ambiguous utterances. More critically, existing S2ST systems offer little explicit modeling of source language information at either the acoustic or linguistic level, despite its central role in accurate multilingual translation.

In this paper, we propose \textbf{S2ST-Omni}, a compositional S2ST framework that couples a SpeechLLM-based S2TT frontend with a plug-and-play pretrained TTS backend. Our key idea is to bridge the representation gap between a pretrained Whisper-based speech encoder \cite{radford2023robust} and a Qwen3 LLM \cite{yang2025qwen3} through structured adaptation and hierarchical language-aware augmentation, thereby better exploiting their cross-modal and multilingual capabilities.
First, we introduce a hybrid speech adapter that follows a "local-then-global" strategy: depthwise separable convolutions first aggregate local acoustic patterns, learnable downsampling then shortens the sequence while preserving salient information, and global self-attention finally models long-range semantic dependencies on the compressed features.
Unlike Conformer-style adapters that interleave convolution and attention at every layer \cite{dubey2024llama,xu2025fireredasr}, this hierarchical design adheres to a "denoise-then-relate" principle, allowing global attention to operate on cleaner and more abstract representations with reduced computational cost.
Second, to explicitly exploit source language information, we propose a \textbf{hierarchical language-aware} (HLA) architecture that operates jointly at acoustic and linguistic levels. On the acoustic side, \textbf{Language-Aware Dual-CTC} (LA-Dual-CTC) comprises a source-CTC and a target-CTC branch. The target-CTC is a standard CTC head supervising target-language alignment, whereas the source-CTC applies FiLM-based language conditioning \cite{perez2018film} on intermediate adapter features, using feature-wise affine modulation and a learnable gate to inject language-specific acoustic biases while preserving source-content fidelity. On the linguistic side, \textbf{Language-Aware Prompting} (LAP) dynamically constructs prompts based on the detected source language, explicitly marking the language at the LLM input and triggering language-specific translation behavior to resolve cross-lingual ambiguities. Importantly, language identity is obtained without introducing a dedicated classifier: during training we use ground-truth labels, and at inference we reuse Whisper encoder features and its decoder for language prediction.
Third, to enable efficient optimization of the S2TT frontend, we present a task-specific \textbf{progressive fine-tuning} (PFT) strategy. In Stage~I, we freeze the Whisper encoder and Qwen3, and train only the hybrid adapter with LA-Dual-CTC to first establish reliable speech-text alignment. In Stage~II, we insert LoRA \cite{hu2022lora} adapters into the LLM and continue to update the LA-Dual-CTC-augmented adapter with reduced learning rates, injecting translation capability on top of a converged alignment while mitigating gradient interference between newly initialized and pretrained components. On the synthesis side, \textit{S2ST-Omni} can seamlessly integrate any state-of-the-art (SOTA) TTS model \cite{chen2024takin,peng2025vibevoice,zhou2025indextts2,zhou2025voxcpm}, without retraining the S2TT frontend.

In summary, we introduce \textit{S2ST-Omni}, a compositional S2ST framework that integrates a SpeechLLM-based S2TT frontend with a plug-and-play TTS backend. Our approach bridges Whisper and Qwen3 via a local-then-global hybrid adapter, and injects source-language information through hierarchical language-aware augmentation (LA-Dual-CTC and LAP). To ensure stable optimization, we propose a task-specific progressive fine-tuning strategy that establishes a robust speech--text alignment foundation before refining translation capabilities via LoRA. Experiments on CVSS-C \cite{jia2022cvss} show that \textit{S2ST-Omni} achieves the best BLEU and ASR-BLEU scores among strong E2E and compositional baselines on French, German, and Spanish $\rightarrow$ English translation.

\section{BACKGROUND}
\label{sec:BACKGROUND}
  
\subsection{Speech-to-Speech Translation} 
Existing S2ST systems broadly fall into two paradigms: E2E and compositional.  
E2E models such as Translatotron series ~\cite{jia2019direct,jia2022translatotron} and SeamlessM4T~\cite{barrault2023seamlessm4t} map source speech directly to target speech within a single network, but must jointly learn speech understanding, translation, and generation, which complicates optimization and typically demands large amounts of speech–speech parallel data. 
Compositional approaches instead factorize S2ST into an S2TT frontend plus a TTS backend.  
ComSpeech~\cite{fang2024can} links pretrained S2TT and TTS models via CTC-based vocabulary adapters, while StreamSpeech~\cite{zhang2024streamspeech} builds a streaming cascaded system with online S2TT and expressive TTS.  
Despite their advantages in interpretability and modularity, these systems still rely on relatively shallow adapters and task-specific S2TT training, and they do not explicitly model source-language information at the acoustic or linguistic level, leaving room for more language-aware and SpeechLLM-centric designs.

\subsection{Speech Large Language Models}
Recent speech–LLM frameworks extend text LLMs to audio either via discrete tokens (e.g., SpeechGPT~\cite{zhang2023speechgpt}, AudioPaLM~\cite{rubenstein2023audiopalm}) or continuous adapters that connect speech encoders such as Whisper~\cite{radford2023robust} to LLM backbones (e.g., SALMONN~\cite{tang2024salmonn}, Qwen-Audio~\cite{chu2023qwen}, Qwen2-Audio~\cite{chu2024qwen2}).  
These models achieve strong results on ASR and spoken understanding and can perform speech-to-text translation as part of a multitask setup.  
However, they are not tailored for high-accuracy multilingual S2TT and lack explicit mechanisms to inject source-language information into either the acoustic representations or the prompting interface, which is precisely the gap our HLA-based SpeechLLM framework designed to address.

\subsection{CTC for Speech Processing}

Connectionist Temporal Classification (CTC) \cite{graves2006connectionist} provides a principled framework for sequence-to-sequence learning without explicit alignment supervision, which has been widely adopted as an auxiliary objective to improve speech-text alignment in encoder-decoder models \cite{kim2017joint,watanabe2017hybrid,yang2023hybridformer}. Given an input sequence $\mathbf{x} = (x_1, \ldots, x_T)$ and target sequence $\mathbf{y} = (y_1, \ldots, y_U)$ where $U \leq T$, CTC marginalizes over all valid alignments $\pi$ that collapse to $\mathbf{y}$:
\begin{equation}
  P(\mathbf{y}|\mathbf{x}) = \sum_{\pi \in \mathcal{B}^{-1}(\mathbf{y})} P(\pi|\mathbf{x})
\end{equation}
where $\mathcal{B}$ denotes the collapsing function that removes blanks and repeated tokens.

\section{METHODOLOGY}
\label{sec:METHODOLOGY}

\begin{figure*}[htbp]
\centering
    \includegraphics[height=9cm,width=!]{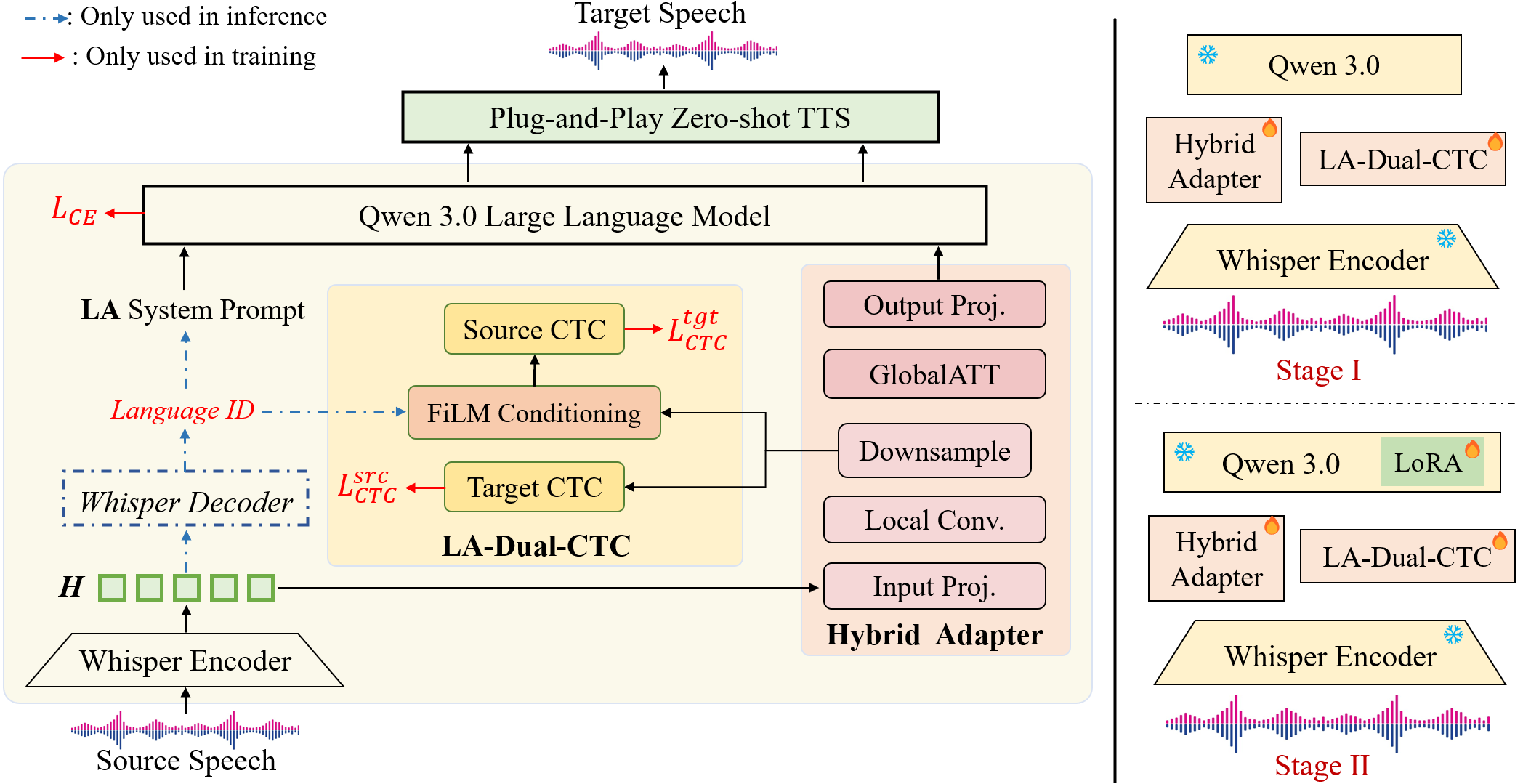}
    \caption{Overall architecture of the proposed S2ST-Omni framework. LA denotes language-aware.}
    \label{fig:s2st-omni}
\end{figure*}

\subsection{System Overview}
\label{sec:overview}

As illustrated in Fig.~\ref{fig:s2st-omni}, \textit{S2ST-Omni} adopts a compositional architecture that decomposes S2ST into two largely independently optimizable modules: 
a high-accuracy S2TT frontend and a flexible TTS backend.
The S2TT frontend comprises four core components:
(i) a Whisper Large-v3 encoder that provides robust multilingual speech representations;
(ii) a Qwen3-4B LLM that serves as the translation backbone;
(iii) a hybrid speech adapter that bridges Whisper and Qwen3 via a hierarchical "local-then-global" strategy; and
(iv) a HLA augmentation architecture consisting of an acoustic-level LA-Dual-CTC and a linguistic-level LAP.
The TTS backend is kept fully modular and can be instantiated with any SOTA synthesizer, without retraining the S2TT frontend.

\subsection{Hybrid Speech Adapter}
\label{sec:hybrid_adapter}

Adapting continuous speech representations to an LLM requires modeling information across multiple scales, from phonetic-level acoustic patterns to sentence-level semantics.
Conformer-style adapters that interleave convolution and attention in every layer offer a general solution; however, early global mixing can dilute fine-grained local cues, and computing self-attention on long speech sequences is computationally expensive.
At the other extreme, MLP-style adapters that apply only per-frame linear projections are parameter-efficient but lack explicit temporal modeling, thereby shifting the burden of capturing short-term phonetic patterns onto the LLM, which is primarily designed for higher-level semantic reasoning.

We therefore propose a \emph{Hybrid Adapter} that follows a ``local-then-global'' hierarchy.
Given Whisper encoder outputs $\mathbf{H} \in \mathbb{R}^{T \times d_s}$ with frame length $T$ and hidden size $d_s = 1280$, we first project them into an adapter hidden space $d_h = 1024$ via a linear layer with LayerNorm and GELU, yielding $\mathbf{H}^{(0)} \in \mathbb{R}^{T \times d_h}$.
On top of $\mathbf{H}^{(0)}$, we stack two depthwise separable convolution blocks.
Each block applies LayerNorm, a depthwise-separable 1D convolution with kernel size $k=7$ to capture local acoustic patterns, followed by a position-wise feed-forward network (FFN) with expansion ratio~4 and a residual connection.

To shorten the sequence before global attention, we apply a strided 1D convolution with kernel size of 5 and stride of 2 to the final output $\mathbf{H}^{(N)}$, followed by LayerNorm and GELU, obtaining a downsampled representation $\mathbf{H}^{\text{down}} \in \mathbb{R}^{T' \times d_h}$ with $T' = \lceil T / 2 \rceil$.
The learnable downsampling adaptively preserves translation-relevant information while discarding redundant frames.
We then feed $\mathbf{H}^{\text{down}}$ into two standard Transformer blocks, each composed of multi-head self-attention with 4 heads and a FFN (expansion ratio~4), both wrapped with LayerNorm and residual connections, to model long-range semantic dependencies and produce a global representation $\mathbf{H}^{\text{glob}} \in \mathbb{R}^{T' \times d_h}$.
Finally, we apply LayerNorm and a linear projection $\mathbf{W}_{\text{out}} \in \mathbb{R}^{d_h \times d_l}$ with $d_l = 3584$ to map $\mathbf{H}^{\text{glob}}$ into the LLM hidden space, yielding $\mathbf{Z} \in \mathbb{R}^{T' \times d_l}$.
This sequence $\mathbf{Z}$ is fed to the LLM, while the intermediate features $\mathbf{H}^{\text{down}}$ are reused by LA-Dual-CTC for language-aware auxiliary supervision.

\subsection{HLA Augmentation Architecture}
\label{sec:dual-lang}

In multilingual S2ST, we suppose that explicit source-language awareness is crucial for accurate translation, as different languages exhibit distinct syntactic structures (e.g., verb-final German clauses, post-nominal French adjectives), word-order patterns, and phonological systems.
To capture these differences, we propose a \emph{HLA augmentation} architecture that injects source-language information at two complementary levels:
(i) LAP operates at the LLM input to guide high-level translation behavior; and
(ii) LA-Dual-CTC operates on adapter features to modulate low-level acoustic representations.
Together, they provide bottom-up and top-down language awareness.

\subsubsection{LAP}
Pretrained LLMs acquire rich cross-lingual knowledge from large-scale multilingual corpora, but effectively leveraging this knowledge requires explicit cues about the input language and task.
LAP provides such cues by dynamically constructing language-aware prompts.

Given a source-language identifier $\ell$ (obtained from Whisper’s built-in language ID during inference, and from dataset labels during training), we instantiate a prompt $\mathcal{P}(\ell)$ using a simple template:
“\emph{The following is \textsc{[LANG]} speech. Translate it accurately into English.}”,  
where \textsc{[LANG]} is the full language name corresponding to $\ell$ (e.g., \texttt{fr} $\rightarrow$ \textit{French}).  
We encode $\mathcal{P}(\ell)$ with the LLM tokenizer to obtain prompt embeddings $\mathbf{E}(\mathcal{P}(\ell))$, which are concatenated with the speech representation $\mathbf{Z}$ and the target text embeddings $\mathbf{E}(\mathbf{y})$ to form the LLM input sequence
\begin{equation}
    \begin{split}
\mathbf{X} = \big[\mathbf{E}(\mathcal{P}(\ell));\, \mathbf{Z};\, \mathbf{E}(\mathbf{y})\big].
    \end{split}
\end{equation}

By this means, LAP is able to explicitly declare the source language and steer the LLM towards the appropriate language-pair translation mode, without introducing additional cost.

\subsubsection{LA-Dual-CTC}

Although LAP operates at the linguistic level, we also aim to encode language-specific inductive biases directly into the speech representation. To this end, we design \emph{LA-Dual-CTC}, which injects source-language information into the CTC heads via FiLM-style conditioning.

Detailed, we apply LA-Dual-CTC to the downsampled intermediate features $\mathbf{H}^{\text{down}}$ from the Hybrid Adapter (Section~\ref{sec:hybrid_adapter}) rather than to its final output. The intuition behind is that this design ensures that $\mathbf{H}^{\text{down}}$ both preserves the monotonic alignment structure required by CTC and retains sufficient acoustic detail, while decoupling the CTC objective from the subsequent global-attention layers that are primarily responsible for high-level semantic modeling.
Then, we obtain a language embedding $\mathbf{e}_\ell = \mathrm{Embed}(\ell)$ and feed it into a small MLP to generate FiLM parameters $(\boldsymbol{\gamma}_\ell, \boldsymbol{\beta}_\ell) \in \mathbb{R}^{d_h} \times \mathbb{R}^{d_h}$, where $\ell$ denotes the source language.
To control the overall strength of language conditioning, we introduce a learnable scalar gate $g$.
Empirically, we initialize $g = 0.5$, which provides a moderate degree of modulation at the start of training and allows the model to adaptively increase or decrease its reliance on language-specific conditioning as learning progresses.
The gated FiLM transformation produces language-conditioned features
\begin{equation}
    \begin{split}
\mathbf{Z}^{\text{src}} = (1 + g\,\boldsymbol{\gamma}_\ell) \odot \mathbf{H}^{\text{down}} + g\,\boldsymbol{\beta}_\ell,
    \end{split}
\end{equation}
on which the \emph{source} CTC head is applied to produce logits $\mathbf{O}^{\text{src}}$.
Because the target language is fixed to English, the \emph{target} CTC head operates directly on $\mathbf{H}^{\text{down}}$ without language conditioning to produce logits $\mathbf{O}^{\text{tgt}}$.


\subsection{Task-Specific PFT}
\label{sec:progressive_ft}

Optimizing the S2TT frontend jointly over a frozen Whisper encoder, a large pretrained LLM, a newly initialized adapter, and auxiliary CTC heads is challenging due to heterogeneous initialization and optimization dynamics.
We therefore adopt a task-specific \emph{PFT} strategy, summarized in Algorithm~\ref{alg:pft}, which proceeds in two stages:
Stage~I emphasizes building a reliable speech--text alignment under relatively strong CTC supervision, while Stage~II refines translation quality via parameter-efficient LLM adaptation with weaker CTC regularization.

\begin{algorithm}[t]
  \caption{Task-Specific PFT}
  \label{alg:pft}
  \begin{algorithmic}[1]
    \REQUIRE Dataset $\mathcal{D} = \{(x,y,\ell)\}$, encoder parameters $\theta_{\text{enc}}$, LLM parameters $\theta_{\text{llm}}$
    \STATE Initialize Hybrid Adapter $\theta_{\text{adpt}}$, LA-Dual-CTC heads $\theta_{\text{ctc}}$, LoRA parameters $\theta_{\text{lora}}$
    \STATE Freeze $\theta_{\text{enc}}$ and base $\theta_{\text{llm}}$ in all stages
    \STATE \textbf{Stage I (alignment):} optimize $\theta_{\text{adpt}}, \theta_{\text{ctc}}$
    \FOR{$(x,y,\ell)\sim\mathcal{D}$}
      \STATE Compute losses $\mathcal{L}_{\text{CE}}, \mathcal{L}_{\text{CTC}}^{\text{src}}, \mathcal{L}_{\text{CTC}}^{\text{tgt}}$
      \STATE Update $\theta_{\text{adpt}}, \theta_{\text{ctc}}$ w.r.t. $\mathcal{L}^{(1)}$
    \ENDFOR
    \STATE \textbf{Stage II (translation):} enable $\theta_{\text{lora}}$, reduce lr for $\theta_{\text{adpt}}, \theta_{\text{ctc}}$
    \FOR{$(x,y,\ell)\sim\mathcal{D}$}
      \STATE Compute losses $\mathcal{L}_{\text{CE}}, \mathcal{L}_{\text{CTC}}^{\text{src}}, \mathcal{L}_{\text{CTC}}^{\text{tgt}}$
      \STATE Update $\theta_{\text{adpt}}, \theta_{\text{ctc}}, \theta_{\text{lora}}$ w.r.t. $\mathcal{L}^{(2)}$
    \ENDFOR
    \ENSURE Trained S2TT frontend
  \end{algorithmic}
\end{algorithm}

\subsubsection{Stage I: Alignment Foundation}
In Stage~I, we freeze both the Whisper encoder and the base Qwen3 parameters, and train only the Hybrid Adapter and LA-Dual-CTC.
The LLM therefore acts as a fixed conditional language model, providing stable gradients while the adapter learns to produce features that are jointly aligned with source and target text under dual CTC supervision.

Let $\mathcal{L}_{\text{CE}}$ denote the autoregressive cross-entropy loss of the LLM decoder, and let $\mathcal{L}_{\text{CTC}}^{\text{src}}$ and $\mathcal{L}_{\text{CTC}}^{\text{tgt}}$ denote the CTC losses on the source and target branches of LA-Dual-CTC, respectively.
The Stage~I objective is
\begin{equation}
  \mathcal{L}^{(1)}
  = \mathcal{L}_{\text{CE}}
  + \alpha_1\,\mathcal{L}_{\text{CTC}}^{\text{src}}
  + \beta_1\,\mathcal{L}_{\text{CTC}}^{\text{tgt}},
  \label{eq:loss_stage1}
\end{equation}
where we assign relatively strong CTC weights, empirically setting $\alpha_1 = 0.1$ and $\beta_1 = 0.2$ to prioritize robust speech--text alignment.

\subsubsection{Stage II: Translation Enhancement}
In Stage~II, we introduce parameter-efficient adaptation on the LLM while continuing to refine the Hybrid Adapter and LA-Dual-CTC with reduced learning rates.
Specifically, we apply LoRA to the key projection matrices of Qwen3, i.e., the query and value projections in the self-attention modules, with rank $r = 8$, scaling factor $\alpha = 32$, and dropout rate $0.1$.
LoRA parameters are trained with a higher learning rate than the adapter and CTC heads, allowing translation-specific knowledge to be injected into the LLM on top of the converged alignment from Stage~I.

To shift the optimization focus towards E2E translation quality while still maintaining useful alignment signals, we down-weight the CTC losses in this stage and adopt
\begin{equation}
  \mathcal{L}^{(2)}
  = \mathcal{L}_{\text{CE}}
  + \alpha_2\,\mathcal{L}_{\text{CTC}}^{\text{src}}
  + \beta_2\,\mathcal{L}_{\text{CTC}}^{\text{tgt}},
  \label{eq:loss_stage2}
\end{equation}
with $\alpha_2 = 0.01$ and $\beta_2 = 0.05$.

\subsection{TTS Backend}
\label{sec:tts_backend}

S2ST-Omni supports plug-and-play integration with any SOTA speech synthesizer as the TTS backend.
The intermediate English text produced by the S2TT frontend is passed directly to an external TTS model, without any task-specific coupling.
In our experiments, we instantiate the backend with IndexTTS 2~\cite{zhou2025indextts2}, a zero-shot TTS system that can generate natural and fluent speech.
This modular design allows practitioners to select or swap TTS systems according to application requirements without any retraining or modifying the S2TT frontend, reducing development and deployment overhead.

\section{EXPERIMENTS}
\label{sec:EXPERIMENTS}


\begin{table*}[ht]
\centering
\caption{Overall performance comparison among evaluated S2ST methods on CVSS-C.}
\label{tab:speech_translation}
\resizebox{\textwidth}{!}{
\begin{tabular}{cccccccccc}
\toprule
\multirow{2}{*}{\textbf{Models}} &
\multicolumn{2}{c}{\textbf{FR $\rightarrow$ EN}} &
\multicolumn{2}{c}{\textbf{ES $\rightarrow$ EN}} &
\multicolumn{2}{c}{\textbf{DE $\rightarrow$ EN}} &
\multicolumn{2}{c}{\textbf{Avg.}} \\
\cmidrule(lr){2-3} \cmidrule(lr){4-5} \cmidrule(lr){6-7} \cmidrule(lr){8-9}
& \textbf{BLEU}$\uparrow$ & \textbf{ASR-BLEU}$\uparrow$
& \textbf{BLEU}$\uparrow$ & \textbf{ASR-BLEU}$\uparrow$
& \textbf{BLEU}$\uparrow$ & \textbf{ASR-BLEU}$\uparrow$
& \textbf{BLEU}$\uparrow$  & \textbf{ASR-BLEU}$\uparrow$ \\
\midrule
Ground Truth     & -     & 84.52 & -     & 88.54 & -     & 75.53 & -     & -    \\
\midrule
Translatotron    & -     & 16.96 & -     &  8.72 & -     &  1.97 & -     &  9.22 \\
Translatotron2   & 28.82 & 26.07 & 25.82 & 22.93 & 18.66 & 16.91 & 24.43 & 21.97 \\
S2UT             & -     & 22.23 & -     & 18.53 & -     &  2.99 & -     & 14.58 \\
UnitY            & -     & 27.77 & -     & 24.95 & -     & 18.74 & -     & 23.82 \\
DASpeech         & -     & 25.03 & -     & 21.37 & -     & 16.14 & -     & 20.85 \\
ComSpeech        & 30.72 & 28.15 & 26.51 & 24.80 & 19.41 & 18.16 & 25.55 & 23.70 \\
StreamSpeech     & 32.60 & 28.45 & 30.35 & 27.25 & 23.36 & 20.93 & 28.77 & 25.54 \\
Hibiki           & -     & 30.50 & -     &   -   & -     &   -   & -     &  -    \\
Ours             & \textbf{35.83} & \textbf{33.20}
                 & \textbf{37.85} & \textbf{35.90}
                 & \textbf{33.34} & \textbf{31.25}
                 & \textbf{35.67} & \textbf{33.45} \\
\bottomrule
\end{tabular}
}
\end{table*}

\begin{table*}[t]
  \centering
  \caption{Effect of the proposed hybrid speech adapter against Conformer-based and MLP-based adapters.}
  \label{tab:adapter}
  \resizebox{\textwidth}{!}{
  \begin{tabular}{ccccccccc}
    \toprule
    \multirow{2}{*}{\textbf{Model}} &
    \multicolumn{2}{c}{\textbf{Fr $\rightarrow$ En}} &
    \multicolumn{2}{c}{\textbf{Es $\rightarrow$ En}} &
    \multicolumn{2}{c}{\textbf{De $\rightarrow$ En}} &
    \multicolumn{2}{c}{\textbf{Avg.}} \\
    \cmidrule(lr){2-3} \cmidrule(lr){4-5} \cmidrule(lr){6-7} \cmidrule(lr){8-9}
    & \textbf{BLEU}$\uparrow$ & \textbf{ASR-BLEU}$\uparrow$
    & \textbf{BLEU}$\uparrow$ & \textbf{ASR-BLEU}$\uparrow$
    & \textbf{BLEU}$\uparrow$ & \textbf{ASR-BLEU}$\uparrow$
    & \textbf{BLEU}$\uparrow$ & \textbf{ASR-BLEU}$\uparrow$ \\
    \midrule
    Ours  & \textbf{35.83} & \textbf{33.20} & \textbf{37.85} & \textbf{35.90} & \textbf{33.34} & \textbf{31.25} & \textbf{35.67} & \textbf{33.45} \\
    w/ Conformer & 34.60 & 31.49 & 36.78 & 34.64 & 32.69 & 30.34 & 34.69 & 32.16 \\
    w/ MLP     & 32.30 & 28.96 & 34.70 & 32.47 & 30.09 & 27.54 & 32.37 & 29.66 \\
    \bottomrule
  \end{tabular}
  }
\end{table*}

\subsection{Experimental Settings}

\subsubsection{Dataset}
We evaluate S2ST-Omni on the CVSS-C benchmark~\cite{jia2022cvss}, which provides parallel triplets $\langle$source speech, target text, target speech$\rangle$ for 21 source languages translated into English.  
Following the protocol of~\cite{fang2024can,zhang2024streamspeech} for fair comparison, we evaluate S2ST-Omni on three representative directions: French$\rightarrow$English (Fr$\rightarrow$En), Spanish$\rightarrow$English (Es$\rightarrow$En), and German$\rightarrow$English (De$\rightarrow$En).  
Source speech is resampled to 16~kHz, and target speech to 22.05~kHz.

\subsubsection{Implementation Details}
We use Whisper Large-v3~\cite{radford2023robust} as a frozen speech encoder and Qwen3-4B~\cite{yang2025qwen3} as the LLM backbone.  
For LA-Dual-CTC, we employ 8k and 4k SentencePiece vocabularies for the multilingual source and English target branches, respectively, both trained on the source and target text of our training set.

Training is implemented in PyTorch on two NVIDIA A6000 GPUs.  
In Stage~I, we train only the Hybrid Adapter and LA-Dual-CTC for 150k steps using AdamW with a cosine schedule, learning rates of $1\times10^{-5}$ (adapter) and $5\times10^{-5}$ (LA-Dual-CTC), and 1k warmup steps.  
In Stage~II, we attach LoRA modules and continue training for another 150k steps with learning rates $5\times10^{-6}$ (adapter), $1\times10^{-6}$ (LA-Dual-CTC), and $5\times10^{-5}$ (LoRA).  
We use a batch size of 4 per GPU with gradient accumulation of 4.

\subsubsection{Evaluation}
We compare S2ST-Omni against eight strong S2ST systems: Translatotron / Translatotron~2~\cite{jia2019direct,jia2022translatotron}, S2UT~\cite{lee2022direct}, DASpeech~\cite{fang2023daspeech}, UnitY~\cite{inaguma2023unity}, ComSpeech~\cite{fang2024can}, StreamSpeech~\cite{zhang2024streamspeech}, and Hibiki~\cite{pmlr-v267-labiausse25a}.

Translation quality is assessed using BLEU and ASR-BLEU.  
For ASR-BLEU, generated speech is first transcribed by a pretrained wav2vec~2.0 ASR model,\footnote{\url{https://dl.fbaipublicfiles.com/fairseq/wav2vec/wav2vec_vox_960h_pl.pt}} and BLEU~\cite{papineni2002bleu} is then computed with SacreBLEU.\footnote{\url{https://github.com/mjpost/sacrebleu}}

\subsection{Main Results}

As shown in Table~\ref{tab:speech_translation}, \textit{S2ST-Omni} consistently outperforms all evaluated S2ST systems on CVSS-C across the three translation directions.
Compared with the strongest compositional baseline StreamSpeech, our model improves the average BLEU score from 28.77 to 35.67 and the average ASR-BLEU from 25.54 to 33.45, corresponding to relative gains of roughly $+24\%$ BLEU and $+31\%$ ASR-BLEU.
These results confirm that the SpeechLLM-based S2TT frontend, together with our LA-Dual-CTC augmented hybrid adapter, HLA architecture, and task-specific PFT, yields a substantially stronger and more robust S2ST system while preserving a fully modular TTS backend.

Breaking down the results by language pair, \textit{S2ST-Omni} achieves consistent improvements.
On the relatively challenging \textbf{De$\rightarrow$En} track, our system attains 33.34 BLEU and 31.25 ASR-BLEU, compared with 23.36 and 20.93 for StreamSpeech, yielding large relative gains of about $+43\%$ BLEU and $+49\%$ ASR-BLEU.
For \textbf{Es$\rightarrow$En}, \textit{S2ST-Omni} reaches 37.85 BLEU and 35.90 ASR-BLEU, improving over StreamSpeech (30.35 / 27.25) by approximately $+25\%$ and $+32\%$, respectively.
On \textbf{Fr$\rightarrow$En}, our model obtains 35.83 BLEU and 33.20 ASR-BLEU, which corresponds to relative gains of about $+10\%$ BLEU and $+17\%$ ASR-BLEU over StreamSpeech, and it also surpasses the recently proposed Hibiki that is specifically optimized for ASR-BLEU on this language pair.

Overall, we attribute these consistent gains to the combination of the LA-Dual-CTC–augmented hybrid adapter and the hierarchical language-aware architecture.
The hybrid adapter provides a stronger speech-to-LLM interface by first consolidating local acoustic structure and then modeling long-range dependencies on a downsampled sequence, while LA-Dual-CTC and LAP inject complementary acoustic- and linguistic-level language cues.
Together with the progressive fine-tuning schedule, these components enable \textit{S2ST-Omni} to more effectively exploit Whisper and Qwen3 for multilingual S2TT and S2ST, yielding state-of-the-art performance on CVSS-C.

\begin{table*}[t]
  \centering
  \caption{Effect of the proposed HLA architecture.}
  \label{tab:dual-la}
  \resizebox{\textwidth}{!}{
  \begin{tabular}{ccccccccc}
    \toprule
    \multirow{2}{*}{\textbf{Model}} &
    \multicolumn{2}{c}{\textbf{Fr $\rightarrow$ En}} &
    \multicolumn{2}{c}{\textbf{Es $\rightarrow$ En}} &
    \multicolumn{2}{c}{\textbf{De $\rightarrow$ En}} &
    \multicolumn{2}{c}{\textbf{Avg.}} \\
    \cmidrule(lr){2-3} \cmidrule(lr){4-5} \cmidrule(lr){6-7} \cmidrule(lr){8-9}
    & \textbf{BLEU}$\uparrow$ & \textbf{ASR-BLEU}$\uparrow$
    & \textbf{BLEU}$\uparrow$ & \textbf{ASR-BLEU}$\uparrow$
    & \textbf{BLEU}$\uparrow$ & \textbf{ASR-BLEU}$\uparrow$
    & \textbf{BLEU}$\uparrow$ & \textbf{ASR-BLEU}$\uparrow$ \\
    \midrule
    Ours  & \textbf{35.83} & \textbf{33.20} & \textbf{37.85} & \textbf{35.90} & \textbf{33.34} & \textbf{31.25} & \textbf{35.67} & \textbf{33.45} \\
    w/o LAP              & 35.25 & 32.36 & 37.37 & 35.30 & 33.13 & 31.01 & 35.25 & 32.89 \\
    w/o LA-Dual-CTC      & 35.16 & 32.11 & 37.22 & 35.09 & 33.10 & 30.97 & 35.16 & 32.72 \\
    w/o HLA       & 33.49 & 30.26 & 35.96 & 33.77 & 31.57 & 28.82 & 33.68 & 30.95 \\
    \bottomrule
  \end{tabular}
}
\end{table*}

\begin{table*}[t]
  \centering
  \caption{Effect of the task-specific PFT strategy. ``w/o PFT-II'' denotes removing Stage~II LoRA-based LLM adaptation; ``w/o PFT'' represents training in a single stage without the proposed two-stage schedule.}
  \label{tab:pft}
  \resizebox{\textwidth}{!}{
  \begin{tabular}{ccccccccc}
    \toprule
    \multirow{2}{*}{\textbf{Model}} &
    \multicolumn{2}{c}{\textbf{Fr $\rightarrow$ En}} &
    \multicolumn{2}{c}{\textbf{Es $\rightarrow$ En}} &
    \multicolumn{2}{c}{\textbf{De $\rightarrow$ En}} &
    \multicolumn{2}{c}{\textbf{Avg.}} \\
    \cmidrule(lr){2-3} \cmidrule(lr){4-5} \cmidrule(lr){6-7} \cmidrule(lr){8-9}
    & \textbf{BLEU}$\uparrow$ & \textbf{ASR-BLEU}$\uparrow$
    & \textbf{BLEU}$\uparrow$ & \textbf{ASR-BLEU}$\uparrow$
    & \textbf{BLEU}$\uparrow$ & \textbf{ASR-BLEU}$\uparrow$
    & \textbf{BLEU}$\uparrow$ & \textbf{ASR-BLEU}$\uparrow$ \\
    \midrule
    Ours  & \textbf{35.83} & \textbf{33.20} & \textbf{37.85} & \textbf{35.90} & \textbf{33.34} & \textbf{31.25} & \textbf{35.67} & \textbf{33.45} \\
    w/o PFT-II           & 33.17 & 29.78 & 35.58 & 33.36 & 31.23 & 28.74 & 33.33 & 30.63 \\
    w/o PFT              & 34.24 & 31.12 & 36.41 & 34.24 & 32.46 & 30.08 & 34.37 & 31.81 \\
    \bottomrule
  \end{tabular}
  }
\end{table*}

\subsection{Ablation Study}


\subsubsection{Effect of Hybrid Speech Adapter.}
Table~\ref{tab:adapter} compares our hybrid adapter with Conformer- and MLP-based variants. 
Replacing the hybrid adapter with the Conformer design leads to a consistent performance drop across all directions: the average BLEU score decreases from 35.67 to 34.69 (about $2.8\%$ relative), and the average ASR-BLEU from 33.45 to 32.16 (about $3.9\%$), with similar trends on each language pair. In contrast, the MLP adapter causes substantially larger degradation, reducing the average BLEU/ASR-BLEU to 32.37/29.66, which corresponds to roughly $9\%$ and $11\%$ relative drops, respectively. 
These trends indicate that MLP-style adaptation is insufficient to bridge Whisper and Qwen3 for S2ST, and that densely interleaving convolution and attention (Conformer-style) is also suboptimal, showcasing the effectiveness of the proposed hybrid adapter.

\subsubsection{Effect of HLA Architecture.}
Table~\ref{tab:dual-la} investigates the contribution of our HLA architecture. Removing either LAP or LA-Dual-CTC alone leads to small but consistent performance drops: w/o LAP reduces the average BLEU/ASR-BLEU from 35.67/33.45 to 35.25/32.89, while w/o LA-Dual-CTC yields 35.16/32.72.
In contrast, removing both components (w/o HLA) causes a much larger degradation, bringing the averages down to 33.68 BLEU and 30.95 ASR-BLEU, which corresponds to roughly $5.6\%$ and $7.5\%$ relative drops.
These trends indicate that linguistic-level prompting (LAP) and acoustic-level FiLM conditioning (LA-Dual-CTC) are complementary: each provides modest gains on its own, but jointly encoding source-language information at both the acoustic and linguistic levels yields a markedly stronger inductive bias for multilingual S2ST.

\begin{table}[t]
\centering
\caption{Effect of different TTS backends on the ASR-BLEU metric. }
\label{tab:tts}
\setlength{\tabcolsep}{3pt} 
\resizebox{\columnwidth}{!}{%
\begin{tabular}{lcccc}
\toprule
\textbf{Models} & \textbf{Fr$\rightarrow$En} & \textbf{Es$\rightarrow$En} & \textbf{De$\rightarrow$En} & \textbf{Avg.} \\
\midrule
IndexTTS2  & 33.20 & 35.90 & 31.25 & 33.45 \\
FireRedTTS2  & 31.55 & 35.18 & 30.94 & 32.56 \\
CosyVoice3   & 31.56 & 35.15 & 30.36 & 32.36 \\
ZipVoice  & 31.48 & 35.08 & 30.46 & 32.34 \\
VoxCPM1.5  & 31.35 & 34.82 & 30.20 & 32.09 \\
\bottomrule
\end{tabular}%
}
\end{table}

\subsubsection{Effect of Task-Specific PFT.}
Table~\ref{tab:pft} evaluates the effect of the proposed two-stage PFT strategy.
The Stage-I-only variant (w/o PFT-II), which omits LoRA-based LLM adaptation, exhibits the largest degradation: the average BLEU/ASR-BLEU drops from 35.67/33.45 to 33.33/30.63, with particularly pronounced degradation on De$\rightarrow$En (33.34/31.25 $\rightarrow$ 31.23/28.74).
Collapsing the two phases into a single-stage training (w/o PFT) mitigates this effect but still lags behind the full schedule, reducing the average BLEU/ASR-BLEU to 34.37/31.81 (about $3.6\%$ and $4.9\%$ relative drops).
These results suggest that learning alignment and translation simultaneously introduces detrimental gradient interference, whereas the proposed two-stage schedule—first emphasizing dual-CTC alignment, then refining translation with LoRA-based LLM adaptation—provides a more stable and effective optimization pathway for the S2TT frontend.


\subsubsection{Effect of Various TTS Backends.}
Table~\ref{tab:tts} examines how the choice of TTS backend influences end-to-end performance when the S2TT frontend of \textit{S2ST-Omni} is kept fixed and only the TTS module is swapped. Overall, the ASR-BLEU scores are relatively stable across five heterogeneous TTS systems: the average gap between the best (IndexTTS2, 33.45) and the weakest configuration (VoxCPM1.5, 32.09) is within roughly $1.4$ points (about $4\%$ relative), and all backends fall into a narrow band on each language pair.

These results suggest that once a strong SpeechLLM-based S2TT frontend is in place, end-to-end performance is primarily governed by the S2TT component, while different modern TTS backends mainly induce second-order variations related to intelligibility and pronunciation. In practice, this means that \textit{S2ST-Omni} can be paired with a wide range of off-the-shelf TTS systems---selected according to latency, speaker style, or deployment constraints---without requiring any redesign or retraining of the translation module.

\section{Conclusion}
\label{sec:conclusion}

We presented \textit{S2ST-Omni}, a hierarchical language-aware framework for accurate and flexible multilingual speech-to-speech translation. By coupling a SpeechLLM-based S2TT frontend with a plug-and-play TTS backend, S2ST-Omni cleanly decouples translation from synthesis, enabling independent optimization of each module while preserving an interpretable text intermediate. At the core of the frontend, a hybrid "local-then-global'' speech adapter, augmented with LA-Dual-CTC and LAP, provides a strong speechLLM interface that jointly models local acoustic patterns, long-range semantics, and explicit source-language cues. A task-specific two-stage PFT strategy further stabilizes training by first establishing robust dual-CTC alignment and then injecting translation capability via LoRA-based LLM adaptation. Experiments on CVSS-C demonstrate that S2ST-Omni consistently outperforms strong E2E and compositional S2ST baselines on Fr$\rightarrow$En, Es$\rightarrow$En, and De$\rightarrow$En, achieving new state-of-the-art BLEU and ASR-BLEU scores. Ablation studies confirm the effectiveness of each proposed component. We believe these results highlight the promise of HLA-enhanced SpeechLLMs as a principled foundation for future multilingual S2ST systems.

\vfill\pagebreak
\label{sec:refs}

\bibliographystyle{ieee}
\bibliography{refs.bib}

\end{document}